\documentclass[sigconf, nonacm]{aamas} 

\usepackage{balance} 
\usepackage{array}
\usepackage{alltt}

\title{Model Capacity Frontier: A Development Framework for LLM-Supported Social Simulation}

\author{Patrick Zhao}
\authornote{Both authors contributed equally to this research.}
\affiliation{
  \institution{Simon Fraser University}
  \city{Burnaby}
  \country{Canada}}
\email{patrick\_zhao@sfu.ca}

\author{Yuxuan Li}
\authornotemark[1]
\affiliation{
  \institution{Carnegie Mellon University}
  \city{Pittsburgh}
  \country{USA}}
\email{yuxuanll@andrew.cmu.edu}

\author{Hirokazu Shirado}
\affiliation{
  \institution{Carnegie Mellon University}
  \city{Pittsburgh}
  \country{USA}}
\email{shirado@cmu.edu}

\begin{abstract}
Large language model (LLM) agents are increasingly incorporated into agent-based models (ABMs) on the premise that richer, context-sensitive human behavior can improve empirical fidelity. 
However, LLM-supported ABMs introduce an open-ended design space in which subtle modeling and configuration choices, such as prompt design and environment coupling, can substantially alter outcomes. 
Developing LLM-supported ABMs to simulate human collective behavior therefore requires navigating configuration sensitivity while identifying the limitations of the current model architecture.
In this paper, we introduce the \textit{Model Capacity Frontier} (MCF), a framework that characterizes the empirical performance attainable by a specified architecture across its admissible configurations. 
In multi-objective settings, this capacity is represented by an estimated Pareto frontier obtained through systematic configuration exploration.
We demonstrate MCF in crowd simulation of university commencement events with approximately 12,000 agents, using empirical movement data on temporal occupancy, spatial distribution, and flow direction. 
We first characterize the capacity of a conventional crowd model and then evaluate successive LLM-supported architectural extensions. 
In the process, MCF reveals which dimensions of attainable performance expand and which remain limited, providing empirical targets for refinement and re-evaluation. 
Initial LLM integration improves flow direction but degrades temporal occupancy and spatial distribution; subsequent architectural refinement of the LLM layer improves temporal occupancy and recovers spatial performance, while retaining much of the earlier gain in flow direction.
Our case study shows how MCF guides the systematic architectural development of LLM-supported ABMs in configuration-sensitive, open-ended design spaces.
\end{abstract}


\newcommand{\BibTeX}{\rm B\kern-.05em{\sc i\kern-.025em b}\kern-.08em\TeX}

\begin{document}



\pagestyle{plain} 
\fancyhead{}



\maketitle 

\thispagestyle{plain}

\section{Introduction}

Large language models (LLMs) are being increasingly integrated into agent-based models (ABMs) to represent complex, context-sensitive human behavior across diverse social domains \cite{parkGenerativeAgentsInteractive2023, gaoLargeLanguageModels2023, hortonLargeLanguageModels2023}.
Yet more plausible or expressive agent behavior does not necessarily imply greater empirical fidelity. 
LLM simulations can produce convincing responses while systematically distorting behavioral patterns, including the variation observed in human populations \cite{wang2025large, li2025actions, peng2026digital}. 
Empirical validation is therefore essential when such simulations are used as evidence about human behavior or to inform consequential decisions \cite{gaoTakeCautionUsing2025, hullman2026human}.

\begin{figure}[t]
    \centering
    \includegraphics[width=0.7\linewidth]{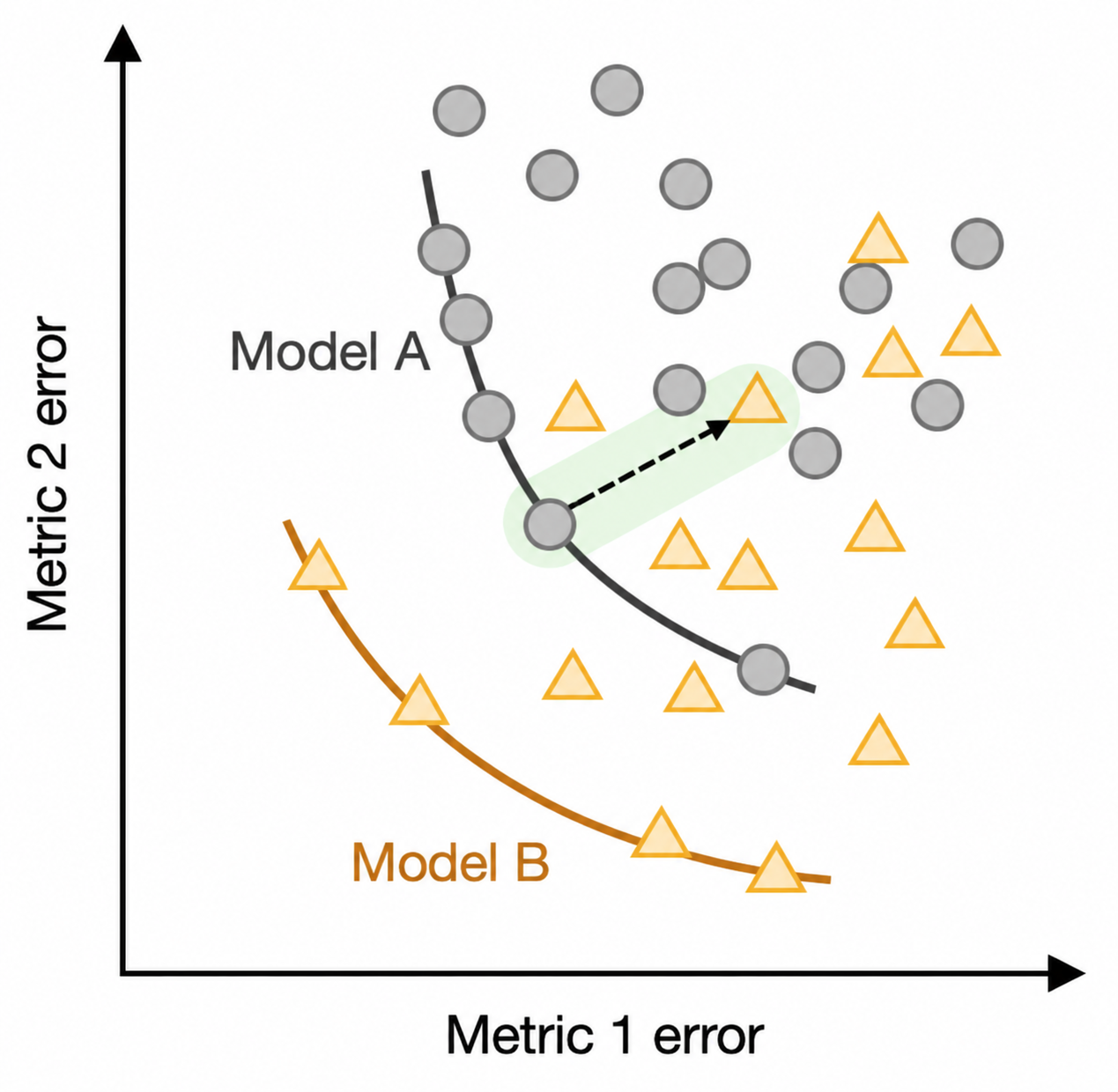}
    \caption{Model Capacity Frontier. Each point represents an admissible configuration from one of two simulation architectures evaluated on two empirical error metrics, and the Pareto frontier traces the nondominated performance attainable by each architecture. Architectural improvement should be judged across configurations, because conclusions based on selected configurations can be fragile, as illustrated here.}
    \label{fig:concept_mcf}
\end{figure}

However, LLM-supported ABMs compound this challenge by enlarging the model-development space: outcomes can depend on prompts, personas, memory, behavioral parameters, and how LLM agents are coupled to the simulated environment \cite{larooijValidationCentralChallenge2025, zhouThisRealLife2024}. 
Parameter calibration and sensitivity are already longstanding challenges in ABMs \cite{shigenakaMASBenchParameterOptimization2021a, banerjee2010validation}, and LLM integration adds open-ended architectural choices. 
When a simulation fails to reproduce an empirical pattern, it can therefore be difficult to tell whether another configuration would suffice or whether the architecture itself lacks necessary modeling capacity \cite{fehlerApproachesResolvingDilemma2006}.
Comparisons between selected configurations can also provide a fragile basis for judging broader modeling progress \cite{hullman2026human}, and may even reverse the apparent conclusion about an architectural extension (Fig. \ref{fig:concept_mcf}).

To address this problem, we introduce the \textit{Model Capacity Frontier} (MCF), a development framework that characterizes the empirical performance attainable by a specified architecture across its admissible configurations. 
In multi-objective settings, MCF is represented by the estimated Pareto frontier of attainable outcomes \cite{deb2011multi}. 
Rather than treating a selected configuration as representative of an architecture, MCF distinguishes configuration-level variation from limitations in the architecture’s attainable empirical performance.

Iterative refinement is often required for ABM development \cite{grimm2005pattern, fehlerApproachesResolvingDilemma2006}. 
MCF structures this process by first characterizing the attainable performance of the current architecture and identifying dimensions or trade-off regions where it remains empirically limited.
Modelers can then introduce a theoretically or empirically motivated architectural refinement and evaluate how the attainable frontier changes.
MCF itself does not prescribe which mechanism should be added; it provides empirical targets for refinement and a common basis for judging whether that refinement expands modeling capacity.
 
We demonstrate this process in a crowd simulation of university commencement events involving over 12,000 attendees. 
We evaluate temporal occupancy, spatial distribution, and flow direction while developing a conventional crowd model through two successive LLM-supported extensions. 
The first improves flow direction but degrades temporal and spatial fit; the second improves temporal occupancy and recovers spatial performance while retaining the earlier flow improvement.

Our main contributions are:
\begin{itemize}
\item \textbf{Model Capacity Frontier (MCF):} a framework for characterizing the empirical performance attainable by a specified architecture across its admissible configurations.
\item \textbf{Frontier-Guided Development:} an iterative workflow that diagnoses whether frontier gaps call for reconfiguration or architectural refinement, and evaluates refinements by the outcomes they newly make attainable.
\item \textbf{Empirical Demonstration:} a 12,000-agent crowd simulation showing how successive LLM-supported extensions reshape attainable performance across multiple empirical dimensions.
\end{itemize}

Although demonstrated in LLM-supported crowd simulation, MCF applies more broadly to configuration-sensitive social simulations with an appropriate empirical evaluation space.
\section{Related Work}
\label{sec:Related Work}

\subsection{Calibration, Validation, and Model Refinement}
\label{subsec:Validation}

Agent-based models are commonly evaluated through parameter calibration, sensitivity analysis, and comparison with empirical patterns \cite{windrum2007empirical, grimm2005pattern}. 
These activities are intertwined with model structure: parameter calibration presupposes an adequate architecture, yet architectural adequacy can only be assessed under particular parameter settings~\cite{fehlerApproachesResolvingDilemma2006}.
Statistical approaches address structural error through model-discrepancy terms~\cite{kennedy2001bayesian} or selection among candidate models~\cite{toni2009approximate}, but typically reduce fit to a likelihood or a scalar distance and assume a fixed set of alternatives.
Prior work has therefore emphasized systematic parameter exploration and empirical validation rather than judging a model from a single realization \cite{shigenakaMASBenchParameterOptimization2021a, banerjee2010validation}.

LLM-supported ABMs intensify this problem because behavior can depend on prompts, personas, memory, model choice, and environment coupling \cite{sumers2023cognitive, gaoTakeCautionUsing2025, larooijValidationCentralChallenge2025}. 
Recent work has accordingly questioned validation based mainly on plausibility, selected demonstrations, or relative improvements between particular LLM configurations \cite{larooijValidationCentralChallenge2025, zhouThisRealLife2024, hullman2026human, zhaoMechanismPlausibilityGenerative2026}. 
This leaves a development question open: when empirical fit remains limited, is the shortcoming addressable through configuration, or does it reflect a limit of the current architecture?
We address this by characterizing empirical capacity across admissible configurations and comparing how it changes across successive architectures.

\subsection{Multi-Objective Evaluation and Pareto Analysis}

Social simulations are often evaluated against multiple empirical targets, and success on one pattern need not imply success on another \cite{grimm2005pattern, lempert2002agent}. 
Scalar calibration can obscure such trade-offs because the resulting optimum depends on how objectives are weighted. 
Multi-objective approaches instead characterize non-dominated configurations through a Pareto frontier \cite{deb2011multi}, and have been applied to ABM calibration across conflicting empirical objectives \cite{rogers2004multi, read2016automated}.

In these applications, Pareto analysis primarily supports parameter calibration or selection within a model architecture. 
We use the same principle for a different development question: what empirical performance combinations are attainable by a specified architecture, and how does that attainable region change after architectural refinement?
Comparing these frontiers across successive architectures provides the basis for the Model Capacity Frontier introduced in Section~\ref{sec:MCF}.

\subsection{Crowd Simulation}

Crowd simulation is a useful setting for studying model capacity because it combines mature mechanistic models with higher-level behavioral decisions and multiple observable outcomes \cite{zheng2009modeling, helbing2015saving}. 
Established approaches such as social-force and velocity-based models reproduce important movement dynamics with relatively parsimonious assumptions \cite{helbingSocialForceModel1995a, tordeuxCollisionFreeSpeedModel2016, seyfriedFundamentalDiagramPedestrian2005}, and have been validated against observed crowd movement and evacuation data \cite{banerjee2010validation, makinoshima2024bayesian}.

Real crowd movement also depends on higher-level decisions about when to move, where to go, and how behavior responds to social and situational context. 
Agent-based crowd models have incorporated group behavior, destination choice, and communication \cite{vizzari2012agent, zhang2022agent}, while recent work has begun integrating language models with crowd and mobility simulation \cite{li2026whatif, yang2025agents, calderon2025cognitive, lammer2026gta}. 
Our case study examines whether successive LLM-supported decision layers expand the empirical performance attainable by a conventional velocity-based model across temporal, spatial, and directional properties of crowd dispersal.
\section{Model Capacity Frontier}
\label{sec:MCF}

MCF has three components: a definition of capacity relative to an explicitly declared model space, a procedure for estimating and comparing frontiers, and diagnostics that turn frontier gaps into development decisions.

\subsection{Model Space and Capacity}
\label{subsec:mcf-def}

A model is specified by an \emph{architecture} $\mathcal{M}$ and a \emph{configuration} $\theta$.
The architecture fixes the model's mechanisms, such as its decision structure, action space, and interaction rules, and for LLM-supported ABMs also the language model, prompts, and the interface through which agents act on the environment.
The configuration collects the quantities that are tuned within that architecture.
Where this boundary is drawn is a modeling choice, and every MCF conclusion holds relative to it.
The choice is especially consequential for LLM-supported ABMs, where prompts and action menus could be placed on either side and where evaluation costs can differ by orders of magnitude across architectures.

Configurations are restricted to an \emph{admissible} space $\Theta_{\mathcal{M}}$, whose ranges are justified by the literature, theory, or measurements that are independent of the evaluation metrics.
Admissibility prevents an architecture from attaining empirical fit through implausible settings, and it is what makes a frontier that presses against a parameter bound informative (Section~\ref{subsec:mcf-dev}).

Each configuration is scored on $k$ empirical targets by the expected error vector $\mathbf{f}(\theta) = \mathbb{E}_{\xi}[\mathbf{e}(\theta,\xi)]$, where $\mathbf{e} \in \mathbb{R}^{k}_{\geq 0}$ is the error of a single run with random seed $\xi$ and lower values indicate better fit.
An error vector $\mathbf{y}$ dominates $\mathbf{y}'$, written $\mathbf{y} \prec \mathbf{y}'$, if $y_i \le y'_i$ for all $i$ and $y_i < y'_i$ for at least one $i$.
The Model Capacity Frontier of $\mathcal{M}$ is the nondominated set of its attainable errors,
\begin{equation}
\mathcal{F}_{\mathcal{M}} = \bigl\{ \mathbf{f}(\theta) \mid \theta \in \Theta_{\mathcal{M}},\ \nexists\, \theta' \in \Theta_{\mathcal{M}} : \mathbf{f}(\theta') \prec \mathbf{f}(\theta) \bigr\}.
\label{eq:frontier}
\end{equation}
The frontier describes the best trade-offs among empirical targets that the architecture can attain within its admissible space, not the capacity of all possible models.

\subsection{Estimating and Comparing Frontiers}
\label{subsec:mcf-est}

Because $\Theta_{\mathcal{M}}$ cannot be evaluated exhaustively, we evaluate a finite sample $\hat{\Theta}_{\mathcal{M}} \subseteq \Theta_{\mathcal{M}}$ and approximate each $\mathbf{f}(\theta)$ by the mean error over $n_s$ seeds, which yields the estimated attainable set $\hat{\mathcal{Y}}_{\mathcal{M}}$.
The estimated frontier $\hat{\mathcal{F}}_{\mathcal{M}}$ is the nondominated subset of $\hat{\mathcal{Y}}_{\mathcal{M}}$.
MCF is agnostic to how $\hat{\Theta}_{\mathcal{M}}$ is constructed, which may rely on low-discrepancy sequences, random sampling, Bayesian optimization, or evolutionary multi-objective search, depending on the computational budget.

The estimate is subject to two errors that act in opposite directions.
Sparse sampling makes $\hat{\mathcal{F}}_{\mathcal{M}}$ conservative, since additional configurations can only move it outward.
Stochastic evaluation makes it optimistic, since configurations that appear nondominated are partly those that drew favorable noise, and this bias grows as $n_s$ decreases.
When architectures are evaluated under different budgets, comparisons should therefore report both the number of configurations and the number of seeds behind each frontier.

We compare architectures through the outcomes that one attains and the other does not.
For an extension $\mathcal{M}_1$ of an architecture $\mathcal{M}_0$, the set of newly attainable outcomes is
\begin{equation}
\mathcal{A}(\mathcal{M}_1 \mid \mathcal{M}_0) = \bigl\{ \mathbf{y} \in \hat{\mathcal{Y}}_{\mathcal{M}_1} \mid \nexists\, \mathbf{y}' \in \hat{\mathcal{Y}}_{\mathcal{M}_0} : \mathbf{y}' \preceq \mathbf{y} \bigr\},
\label{eq:expansion}
\end{equation}
where $\mathbf{y}' \preceq \mathbf{y}$ means $y'_i \le y_i$ for all $i$.
$\mathcal{M}_1$ expands capacity when this set is nonempty, even if it does not dominate the previous frontier everywhere, since an extension may open one trade-off region while receding in another.
Dominance is assessed in all $k$ dimensions, and two-dimensional projections serve only for visualization.
When two architectures share a configuration space, evaluating them at matched configurations additionally reveals how the same underlying parameter setting responds to architectural change.

\subsection{Frontier-Guided Development}
\label{subsec:mcf-dev}

MCF supports an iterative loop in which the modeler estimates $\hat{\mathcal{F}}_{\mathcal{M}_t}$, diagnoses where it falls short of the empirical targets, refines the architecture to $\mathcal{M}_{t+1}$, and evaluates how the attainable frontier changes.
The critical step is diagnosis, because a gap in the frontier does not by itself reveal whether the remedy is another configuration or another architecture.

Evidence for an architectural limitation may come from several sources.
The best configurations may press against an empirically justified parameter bound, the failing dimension may change little across the admissible configuration space, or the behavior required to close the gap may be absent from the architecture's action space.
These patterns are diagnostics rather than proofs, but they indicate when further tuning is unlikely to resolve the observed limitation.

The frontier does not prescribe which mechanism to add.
Refinements should instead be motivated by empirical observation, theory, or domain knowledge rather than by the metric being optimized.
MCF then provides a common basis for evaluating whether a refinement expands the model's empirical capacity.

\section{Crowd Simulation Case}

\begin{figure*}[t]
  \centering
  \includegraphics[width=1.9\columnwidth]{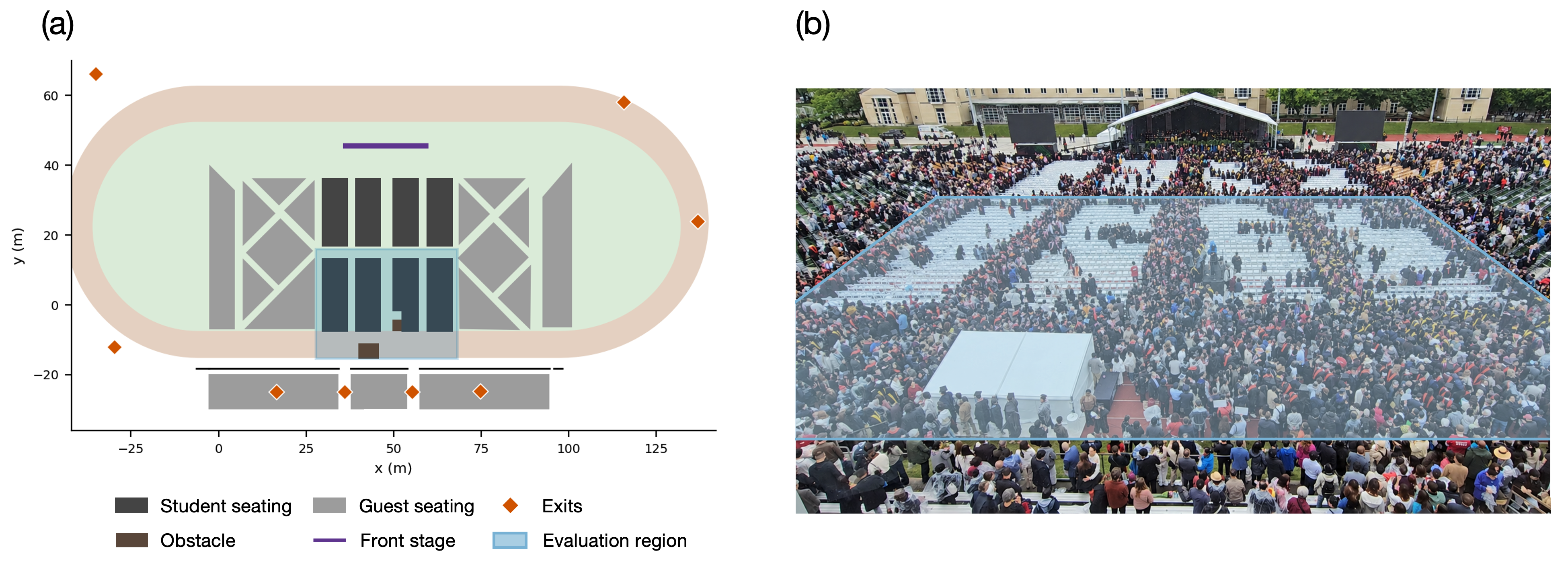}
  \caption{Empirical crowd setting and simulation environment. (a) Top-down simulation environment of the commencement venue, including student and guest seating, stage, obstacles, exits, and the evaluation region. (b) A snapshot from the 2026 commencement recording showing the corresponding real-world scene. The shaded polygon marks the same evaluation region shown in (a), projected onto the camera view for empirical comparison.}
  \label{fig:arena}
\end{figure*}

\subsection{Empirical Context and Validation Data}
\label{subsec:Data}
Our case study models crowd dispersal following a university commencement ceremony held in a football stadium. 
The event contains about 12,300 attendees: 3,000 students seated on the field and 9,300 guests distributed across field-level and grandstand seating (Fig. \ref{fig:arena}a). 
Following the ceremony, attendees exhibit heterogeneous behaviors, including leaving the venue through an exit, remaining near the field, gathering with others, and moving among different destinations. 
These behaviors combine local pedestrian dynamics with higher-level decisions about when and where to move.

We record the event in 2025 and 2026 from a press-box camera overlooking the stadium (Fig. \ref{fig:arena}b). 
The two years provide repeated observations of the same recurring event under the largely similar venue layouts, but with different attendees and realized crowd behavior.
These observations allow us to evaluate whether simulation configurations reproduce crowd patterns consistently across event instances rather than matching the idiosyncrasies or measurement noise of a single realization.

From the videos, we derive both the venue layout and the observed crowd distribution.
We manually identify seating, aisles, stage, obstacles, stairways, and exits, and use landmark correspondences to map the camera view into the top-down simulation coordinate system through a homography.
We then estimate pedestrian positions every 10 seconds over a 25-minute period and map them into the same coordinate system, producing video-derived crowd density measurements that can be compared directly with the simulation.
Details of the spatial calibration and pedestrian-detection pipeline, including validation against manually labeled frames, are provided in Appendix~\ref{app:calibration} and Appendix~\ref{app:measurement}\footnote{Appendices are provided in the supplementary material.}.

Because detection reliability decreases with distance from the camera and in dense regions, we restrict quantitative evaluation to the blue-shaded evaluation region in Fig.~\ref{fig:arena}, while still simulating the full stadium.
This region is well observed by the camera and captures a major intersection through which many attendees pass during dispersal.

\subsection{Empirical Evaluation Metrics}
\label{subsec:Evaluation}

Following Section \ref{sec:MCF}, each configuration $\theta$ is evaluated along three empirical dimensions: temporal occupancy, spatial distribution, and flow direction,
\[
\mathbf{f}(\theta) =[ f_{\mathrm{temporal}}(\theta), f_{\mathrm{spatial}}(\theta), f_{\mathrm{flow}}(\theta)]
\]
Each component measures error, so lower values indicate better performance on all three dimensions.
Each metric is computed separately for the 2025 and 2026 observations using the video-derived density $\rho_{\mathrm{obs}}(\mathbf{x},t)$ and simulated crowd density $\rho_{\theta}(\mathbf{x},t)$ within the evaluation region.
To require configurations to reproduce the recurring crowd pattern across both event instances, we aggregate the two years using the larger error:
\[
f_i(\theta)
=
\max(f_i^{2025}(\theta), f_i^{2026}(\theta)).
\]
Thus, a configuration receives a low error only when it performs well in both years.

\subsubsection{Temporal Occupancy}
The temporal occupancy error measures how accurately the simulation reproduces the number of people present in the evaluation region $R$ over time. 
Let
\[
  N(t)=\int_{R}\rho(\mathbf{x},t)\,\mathrm{d}\mathbf{x},
\]
denote the number of people in $R$ at time $t$.
We define temporal error as the mean absolute difference between simulated and observed occupancy:
\[
  f_{\mathrm{temporal}}(\theta) = \frac{1}{|T|}\sum_{t\in T}
  |N_{\theta}(t) - N_{\mathrm{obs}}(t)| .
\]
\subsubsection{Spatial Distribution}
The spatial distribution error measures how accurately the simulation reproduces \emph{where} people are, independently of the total number present.
For both the observed data and the simulation, we pool density across sampling times and normalize it to unit mass within $R$:
\[
  p(\mathbf{x}) = \frac{\sum_{t\in T} \rho(\mathbf{x},t)}{\sum_{t\in T} N(t)}, \qquad \mathbf{x}\in R
\]
Because $p$ is normalized, this metric captures spatial distribution separately from temporal occupancy.
Crowded moments contribute proportionally more mass than sparse moments.

We measure spatial error using the 1-Wasserstein distance \cite{villaniWassersteinDistances2009},
\[
f_{\mathrm{spatial}}(\theta)
=
W_1
\left(
p_{\theta},
p_{\mathrm{obs}}
\right).
\]
Because the ground distance is measured in meters, $f_{\mathrm{spatial}}$ can be interpreted as the average distance by which simulated crowd mass must be moved to match the observed spatial distribution.

\subsubsection{Flow Direction}
The flow error measures how accurately the simulation reproduces the direction of crowd movement.
We divide the evaluation region into 2\,m cells $c$ and, within each cell, average all walking motion faster than 0.3\,m/s over the evaluation period to obtain a unit direction vector $\mathbf{u}(c)$ (Appendix~\ref{app:measurement}).
We then compute the cosine distance between the simulated and observed directions, weighted by the observed pedestrian mass $m_{\mathrm{obs}}(c)$ in each cell:
\[
  f_{\mathrm{flow}}(\theta) =
  \frac{\sum_{c\in C} m_{\mathrm{obs}}(c)\,
        \bigl(1-\mathbf{u}_{\theta}(c)\cdot\mathbf{u}_{\mathrm{obs}}(c)\bigr)}
       {\sum_{c\in C} m_{\mathrm{obs}}(c)},
\]
where $C$ contains cells with sufficient walking motion in both the simulation and observation.
The error is 0 for aligned directions, 1 for perpendicular directions, and 2 for opposite directions.

\subsection{Shared Simulation Setup}
\label{subsec:Setup}

All model architectures share the same environment, population initialization, pedestrian movement model, and admissible parameter space.
They differ only in the higher-level policy governing when and where agents move.

\subsubsection{Environment and Population}
The simulation represents the real stadium in metric coordinates (Fig.~\ref{fig:arena}), using the spatial calibration described in Appendix~\ref{app:calibration}.
The environment includes the field, track, seating, stage, obstacles, grandstand stairways, and exits.
We simulate 12,278 agents: 2,928 students and 9,350 guests.
Agents begin in their corresponding seating areas, except for the observed number already on the track at the start of the evaluation period (276 in 2025 and 538 in 2026).
Because social-group membership cannot be recovered from the recordings, we synthetically assign each student a party of 1--6 guests seated in the same guest section.


\subsubsection{Routing and Pedestrian Movement}
Given a destination, agents follow a shortest-time route around seating and obstacles (Appendix~\ref{app:pathfinding}).
Local movement follows a collision-free speed model~\cite{tordeuxCollisionFreeSpeedModel2016, chraibiJuPedSim2026}, with heterogeneous desired walking speeds and shared parameters governing time gap, standstill spacing, and neighbor steering.

We define admissible parameter ranges using prior pedestrian research and measurements from the observed events (Table~\ref{tab:theta}; Appendix~\ref{app:config}).
Within each configuration,
\[
v_i^0 \sim \mathcal{N}(\bar v^0,s_{v^0}^2),
\]
while $\mathcal{T}$, $\ell$, $a$, and $D$ are shared across agents.
Configurations are sampled from these ranges using a Sobol low-discrepancy sequence~\cite{sobolDistributionPointsCube1967}.



\begin{table}[t]
  \caption{Admissible parameter ranges used in configuration exploration.}
  \label{tab:theta}
  \small
  \begin{tabular}{llc}
    \toprule
    symbol & meaning & range \\
    \midrule
    $\sigma_d$      & leaving-delay scale (s)& 0--182 \\
    $\bar v^{0}$    & mean desired speed (m/s)      & 1.0--1.6 \\
    $s_{v^{0}}$     & desired-speed spread (m/s)     & 0.2--0.4 \\
    $\mathcal{T}$   & time gap (s)       & 0.8--1.2 \\
    $\ell$          & standstill spacing (m)             & 0.3--0.5 \\
    $a$             & steering strength    & 0--16 \\
    $D$             & steering range (m)   & 0.05--0.3 \\
    \bottomrule
  \end{tabular}
\end{table}
\section{Frontier-Guided Development}
\label{sec:development}

We apply the development loop of Section~\ref{sec:MCF} to the commencement case.
We begin with a conventional crowd model (M0) and then introduce two successive LLM-supported architectures (M1 and M2), both using GPT-6 Luna \cite{openai2026gpt6luna} as the language-model decision component.
For each architecture, we describe its design, estimate its frontier, diagnose the remaining empirical limitation, and ask whether that limitation can plausibly be addressed through another configuration or instead motivates an architectural change.
Table~\ref{tab:architectures} summarizes the behavioral assumptions introduced at each step.

\begin{table}[t]
  \caption{Behavioral assumptions of the three architectures. Bold entries are introduced relative to the preceding architecture.}
  \label{tab:architectures}
  \small
  \setlength{\tabcolsep}{4pt}
  \renewcommand{\arraystretch}{1.1}
  \begin{tabular}{@{}>{\raggedright\arraybackslash}p{1.45cm}
                    >{\raggedright\arraybackslash}p{1.95cm}
                    >{\raggedright\arraybackslash}p{2.0cm}
                    >{\raggedright\arraybackslash}p{2.2cm}@{}}
    \toprule
        & M0 & M1 & M2 \\
    \midrule
    Onset & individual delay & \textbf{group delay} & group delay \\
    Destination & nearest exit & \textbf{LLM-chosen destination} & LLM-chosen destination \\
    Coordination & none & \textbf{group coordination} & \textbf{POI gathering + group coordination} \\
    Stopping & none & \textbf{wait/stage} & \textbf{gather and linger} \\
    Routing & static & \textbf{crowd-aware} & crowd-aware \\
    \bottomrule
  \end{tabular}
\end{table}


All three architectures share the same environment, population, pedestrian movement model, empirical evaluation metrics, and admissible configuration space $\Theta$ described in Section~\ref{subsec:Setup} and Table~\ref{tab:theta}.
Architectural changes are confined to the higher-level decision layer governing when and where agents move; for M1 and M2, the language model, prompts, action space, and query schedule are fixed within each architecture.

We sample configurations from a Sobol low-discrepancy sequence~\cite{sobolDistributionPointsCube1967} over $\Theta$.
Because M0 is inexpensive, we evaluate 512 configurations with three seeds each.
M1 and M2 require LLM calls for agent decisions, so we evaluate the first 16 configurations of the same Sobol sequence with one seed each.
Each M1 or M2 run therefore has a matched M0 configuration with the same pedestrian parameters, which allows us to examine how an architectural change affects the same configuration.
Details of the pilot-based sample-size calculation are provided in Appendix~\ref{app:power}.


Following Section~\ref{subsec:mcf-est}, the sparse M1 and M2 frontiers give conservative evidence wherever they extend beyond the 512-configuration M0 frontier, although their single-seed estimates are more exposed to favorable noise.

\begin{figure*}[t]
    \centering
    \includegraphics[width=0.8\linewidth]{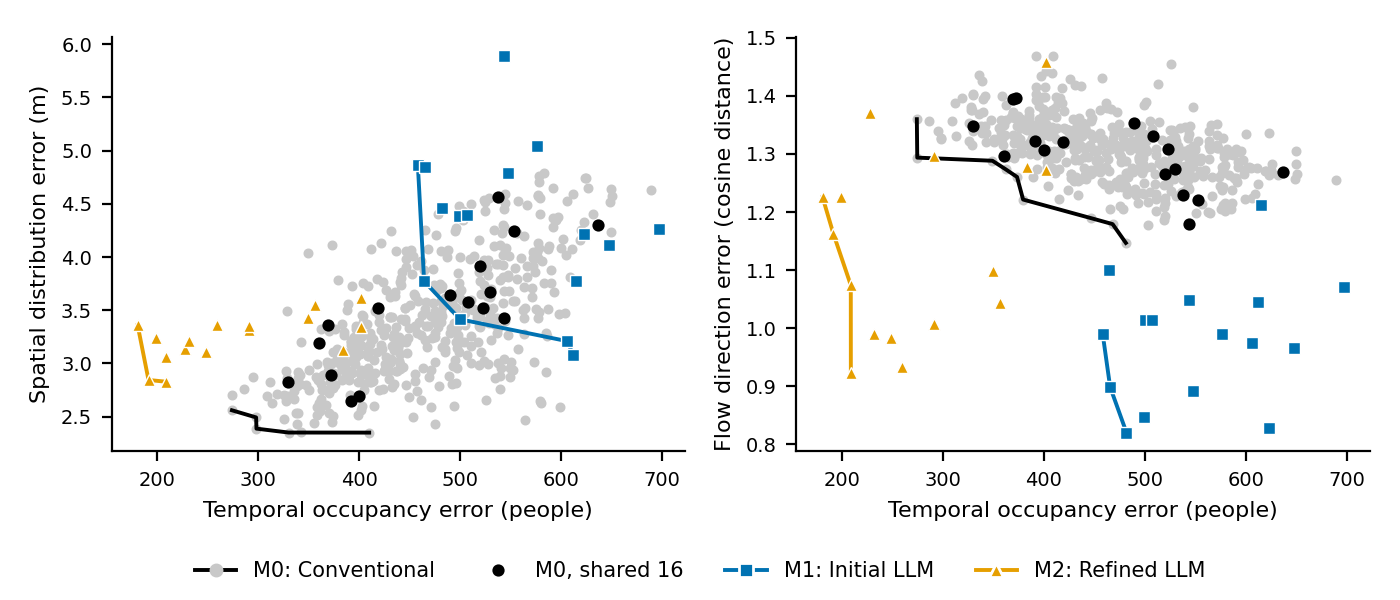}
    \caption{Estimated frontiers in the worse-year view.
    The left panel shows temporal versus spatial error, and the right panel shows temporal versus flow error.
    Lower is better on all axes.
    Gray points show the 512 M0 configurations; black points highlight the 16 M0 configurations at the matched Sobol locations used to evaluate M1 and M2.
    Lines connect the nondominated configurations of each architecture.}
    \label{fig:frontiers}
\end{figure*}

\subsection{M0: Conventional Egress Model}
\label{subsec:m0}

\subsubsection{Design}
In M0, each agent remains seated for a leaving delay drawn from $|\mathcal{N}(0,\sigma_d^2)|$, then walks to its nearest exit and leaves without stopping. 
Agents interact only through the pedestrian movement model.
This follows the standard structure of egress models, combining pre-movement delay~\cite{purserQuantificationBehaviourEngineering2001, lovreglioPreevacuationDatabaseUse2019, kuligowskiReviewBuildingEvacuation2010}, shortest-path exit routing~\cite{kleinmeierVadereOpensourceSimulation2019}, and calibrated pedestrian dynamics~\cite{tordeuxCollisionFreeSpeedModel2016, chraibiJuPedSim2026}.
The admissible range of $\sigma_d$ is constrained by the observed onset of crowd movement: in both years, most attendees leave their seats within about three minutes.
At the upper bound, $\sigma_d = 182$\,s, 90\% of agents begin moving by 300\,s (Appendix~\ref{app:config}).

\subsubsection{Frontier and Diagnosis}
Across M0 configurations, the minimum marginal errors are 274 people for temporal occupancy, 2.35\,m for spatial distribution, and 1.15 for flow direction (Fig.~\ref{fig:frontiers}).
M0 reproduces spatial distribution relatively well, but its temporal and directional behavior reveal limitations that further tuning is unlikely to resolve.

M0 empties the evaluation region too quickly: occupancy falls to half its peak after 7.8\,min, compared with 11.3 and 11.8\,min in the two observations (Fig.~\ref{fig:diagnosis}a).
The best temporal configurations push $\sigma_d$ toward its upper bound, but increasing it further would keep agents seated longer than observed.
The remaining temporal error therefore points to what agents do after they begin moving rather than to insufficient pre-movement delay.

Flow error is even less responsive to configuration, ranging only from 1.15 to 1.47 across all 512 configurations.
Because an error above 1 indicates movement more than orthogonal to the observed direction on average, no explored M0 configuration reproduces the dominant flow well.
This limitation follows from the architecture: under nearest-exit routing, each agent's seat determines its destination, while $\Theta$ only affects movement timing and local walking dynamics.
Observed attendees frequently move toward the field and farther exits, whereas M0 directs them toward the nearest exit (Fig.~\ref{fig:diagnosis}b).
We therefore replace the destination decision policy in M1 while keeping the pedestrian model and $\Theta$ fixed.

\subsection{M1: LLM Decision Layer}
\label{subsec:m1}

\subsubsection{Design.}
Post-ceremony movement is shaped by higher-level social decisions: attendees look for family, coordinate with their party, and choose destinations based on what lies beyond different exits.
Such choices can be represented with rules-based or utility-based policies~\cite{vizzari2012agent, zhang2022agent}, but fitting these policies would require individual choice data that our observations do not provide.
M1 instead replaces nearest-exit routing with an LLM-supported decision layer.

After the inherited leaving delay, each agent chooses among actions such as walking to an exit, going to a family member, or waiting.
Members of a party share messages and can coordinate on an exit.
The pedestrian model and configuration space $\Theta$ remain unchanged from the matched M0 configuration. 
To support the dynamic exit choice, we update the navigation field (see Appendix \ref{app:pathfinding}).
The leaving-delay distribution is also unchanged, although M1 draws one shared delay per family rather than independently per agent to support coordination.
Full prompts and implementation details are provided in the supplementary material.

\subsubsection{Frontier and Diagnosis}
M1 substantially improves directional behavior.
Flow error improves in 15 of 16 matched configurations (median $-0.31$, $d_z=-2.10$), and the best flow error decreases from 1.15 to 0.82 (Table~\ref{tab:transitions}).
Even in the temporal--flow projection, 15 of 16 M1 configurations are not dominated by the M0 frontier, so all 15 belong to $\mathcal{A}(\text{M1} \mid \text{M0})$ in Eq.~\eqref{eq:expansion}.
However, this expansion comes with regressions in the other dimensions: temporal error increases in 15 of 16 matched configurations (median $+81$ people), and spatial error increases in 12 (median $+0.91$\,m).

Simulation replays show that M1 disperses attendees too efficiently.
Although matched M0 and M1 configurations begin moving at the same times, M1 distributes agents across more exits and reduces congestion, causing the evaluation region to empty even faster.
Part of M0's temporal fit was therefore produced by congestion rather than realistic post-ceremony behavior, a compensating error~\cite{grimm2005pattern}.
Observed attendees instead form persistent clusters on the track and near the stage, whereas M1 agents mostly pass through the region.

This limitation is unlikely to be resolved by further tuning of the current configuration space.
The leaving-delay range is already constrained by observed movement onset, and M1's action space lacks a behavior for traveling to a location, gathering there, and remaining before leaving.
Simply asking agents to wait would keep them near their starting locations rather than reproduce the timing and location of the observed clusters.
This motivates M2, which adds explicit gathering behavior to the decision architecture.

\subsection{M2: Gathering at Points of Interest}
\label{subsec:m2}

\subsubsection{Design.}
The M1 diagnosis points to a missing post-ceremony behavior: attendees often travel to a location, gather with others, and remain there before leaving.
In the observations, families meet graduates on the track, take photos near the stage, and wait near exits before departing together.
M2 therefore extends the M1 action space with explicit gathering behavior.
Party members may agree on a point of interest (POI), travel there, wait for the group to assemble, linger, and then coordinate an exit.
Candidate POIs are selected from recurring gathering locations visible in the venue and observations.
All other components are inherited from M1.

\subsubsection{Frontier and Diagnosis}
Adding gathering substantially improves temporal and spatial fit.
Relative to M1, temporal error decreases in all 16 matched configurations (median $-276$ people), and spatial error decreases in 14 (median $-1.18$\,m).
Flow error, however, increases in most configurations (median $+0.19$).
Relative to M0, M2 improves the median error on all three dimensions ($-223$ people, $-0.45$\,m, and $-0.13$), and its temporal--flow frontier extends beyond the M0 frontier.
M2 nevertheless does not recover the lowest spatial error attained by M0.

\begin{figure}[t]
    \centering
    \includegraphics[width=1.0\linewidth]{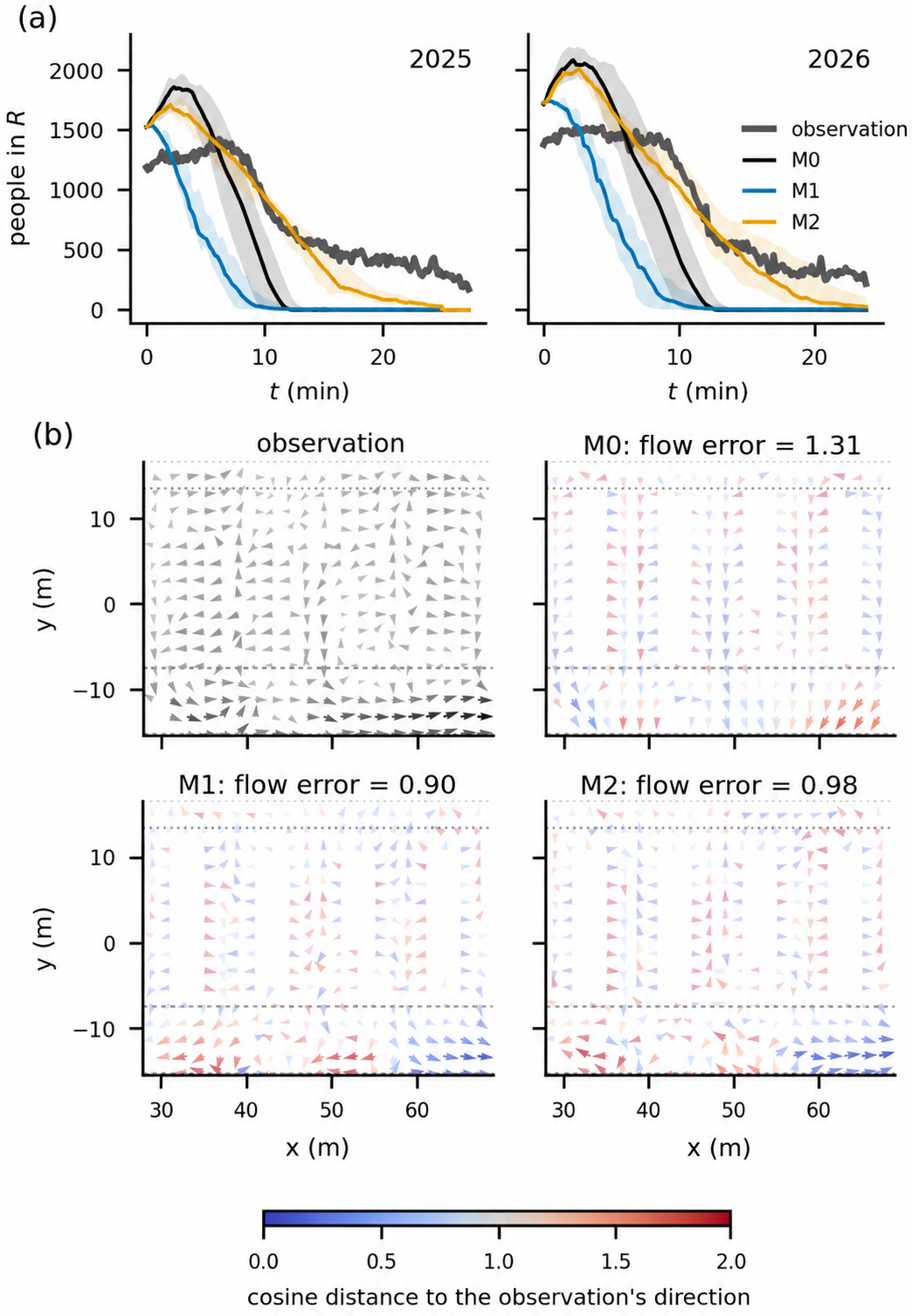}
    \caption{Behavioral diagnosis across architectures.
    (a) Observed occupancy and median simulated occupancy in the evaluation region over time. Shaded areas represent the 25th to 75th percentile over the 16 configurations.
    (b) Mean walking direction by spatial cell in 2026, weighted by observed pedestrian movement. Arrows are scaled by the amount of observed walking.}
    \label{fig:diagnosis}
\end{figure}

The behavioral traces explain this trade-off.
M2 keeps more attendees in the evaluation region and better reproduces the observed occupancy decline (Fig.~\ref{fig:diagnosis}a), consistent with its added gathering behavior.
At the same time, the remaining flow error is concentrated near the track in front of the grandstand (Fig.~\ref{fig:diagnosis}b).
M2 therefore captures the fact that attendees gather and linger, but only approximately where they gather and how they move toward those locations.
These residual errors identify the target for further architectural refinement.

\begin{table}[t]
  \caption{Changes across development over the 16 matched configurations.
  Entries report the median change in error, with the number of configurations that improve in parentheses.
  Negative values indicate better fit and are shown in bold.}
  \label{tab:transitions}
  \setlength{\tabcolsep}{4pt}
  \begin{tabular}{@{}lccc@{}}
    \toprule
    Transition & Temporal & Spatial & Flow \\
    \midrule
    M0 $\rightarrow$ M1
    & $+81$ (1)
    & $+0.91$ (4)
    & $\mathbf{-0.31}$ (15) \\

    M1 $\rightarrow$ M2
    & $\mathbf{-276}$ (16)
    & $\mathbf{-1.18}$ (14)
    & $+0.19$ (3) \\

    M0 $\rightarrow$ M2
    & $\mathbf{-223}$ (15)
    & $\mathbf{-0.45}$ (11)
    & $\mathbf{-0.13}$ (12)\\
    \bottomrule
  \end{tabular}
\end{table}


\section{Discussion}
\label{sec:discussion}

LLM-supported ABMs open a design space in which a disappointing result may stem from the chosen configuration or from the architecture itself.
In our case study, MCF helped distinguish these two sources of error and guided two architectural refinements that moved the attainable frontier into regions the conventional model could not reach.
We now consider what this process reveals beyond the specific case, what it implies for the design of LLM-supported agents, and where its conclusions are limited.

\paragraph{What the frontier revealed.}
First, configuration-level comparisons alone would have suggested a different conclusion about M1.
Across matched configurations, M1 is worse than M0 on two of three metrics in most cases, and a modeler comparing only selected configurations might therefore have abandoned it.
Viewed through the frontier, however, M1 opens a region of flow performance that no M0 configuration attains, and it provides the basis for M2.
The transition from M0 to M1 is thus an empirical instance of the situation sketched in Figure~\ref{fig:concept_mcf}.

Second, an architectural change can expose compensating errors that calibration within a single architecture cannot detect.
M0's temporal fit depended partly on congestion at the nearest exits and on a leaving delay pressed against its bound.
When M1 corrected agents' destinations, these compensations disappeared and the temporal error increased.
A frontier that recedes after a well-motivated change can therefore still be informative, signaling that the predecessor matched the data for the wrong reasons~\cite{grimm2005pattern}.

\begin{figure}[t]
    \centering
    \includegraphics[width=0.8\linewidth]{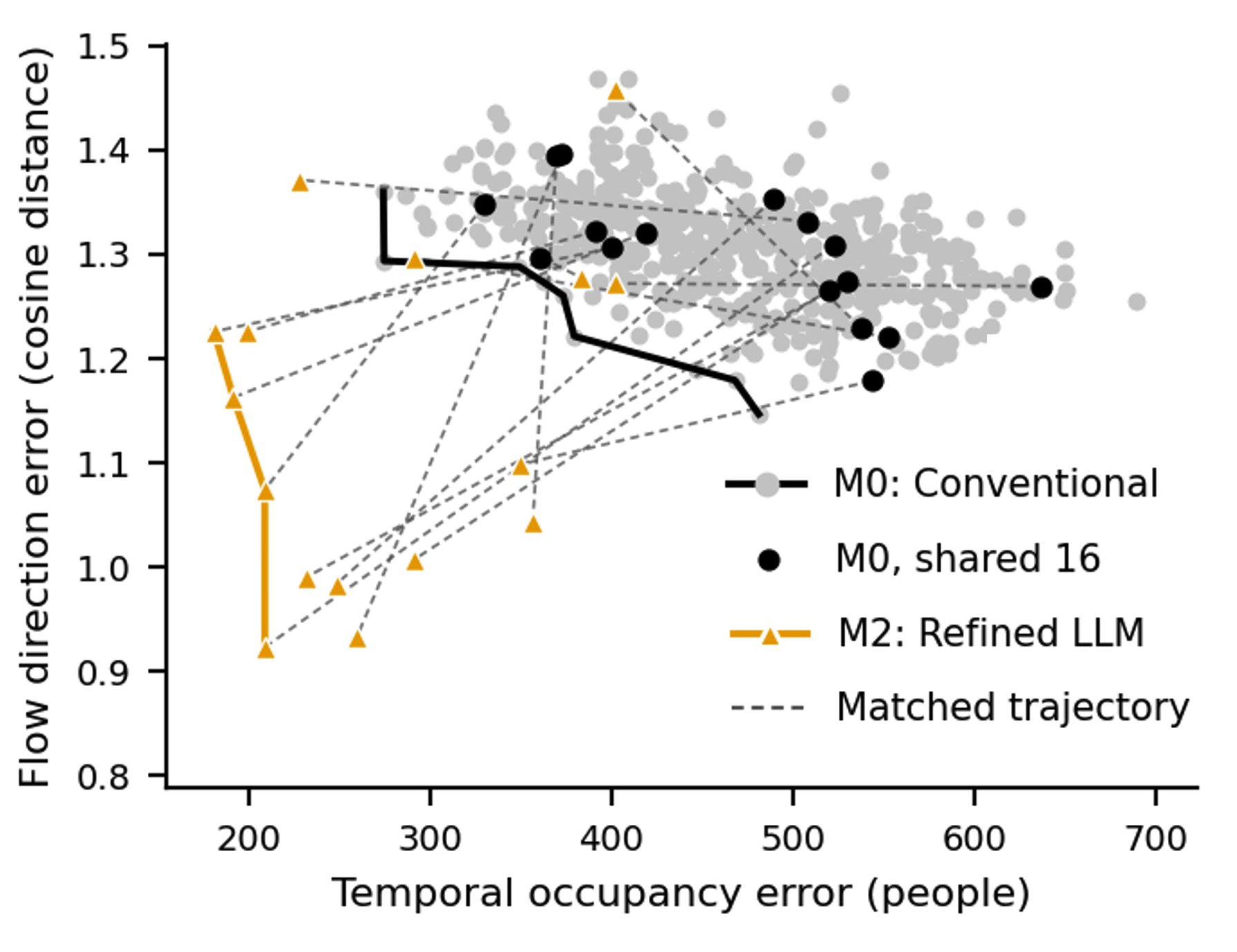}
    \caption{Configuration-matched trajectories from M0 to M2 in the temporal--flow error space.
    Dashed lines connect configurations with identical values of $\Theta$.
    The black and orange points and the corresponding frontier lines are the same as in Figure~\ref{fig:frontiers}.}
    \label{fig:trajectory}
\end{figure}

Third, configuration quality is not preserved across architectures.
The matched trajectories show that configurations performing relatively well under M0 are not necessarily those that perform well under M2, and vice versa (Fig. \ref{fig:trajectory}).
Architectural changes therefore reshape the mapping from configuration space to empirical performance rather than simply shifting all configurations in the same direction.
This makes conclusions based on a single matched comparison fragile: the apparent benefit or cost of an architectural extension can depend strongly on which configuration happens to be selected.

\paragraph{Implications for LLM-supported agents.}
The development trajectory also indicates where an LLM adds modeling capacity.
The LLM layer improved the model where behavior depends on social context and knowledge of the setting, such as choosing destinations and coordinating within a party, while the walker remained the appropriate mechanism for physical movement.
This division of labor favors hybrid architectures in which LLMs govern decisions and established mechanistic models govern motion.

The trajectory further shows that the capacity of an LLM layer is bounded by the action space through which it is coupled to the environment.
M1 agents rarely lingered, because their action space allowed individuals to wait but offered no way for a party to gather, and they defaulted to efficient, goal-directed departure.
This tendency is consistent with evidence that LLM agents behave more purposefully and less variably than the people they represent~\cite{wang2025large, li2025actions, peng2026digital}.
Adding the gathering affordance in M2 changed what the same language model could express.
Evaluating an LLM-supported ABM therefore requires attention to the interface that connects the language model to the simulated world, in addition to the model and its prompts.
MCF makes this interface explicit by treating it as part of the architecture.

\paragraph{Using MCF in practice.}
The MCF protocol carries over to other configuration-sensitive simulations.
However, evaluation dimensions and admissible parameter ranges should be grounded in theory and empirical evidence, because pressure against a parameter bound is interpretable only when that bound is defensible.
Frontier gaps should not automatically be attributed to architectural limitations~\cite{fehlerApproachesResolvingDilemma2006}.
In our case, pressure against empirically justified parameter bounds, weak sensitivity to configuration changes, and missing behaviors in the action space provided practical evidence for distinguishing configuration-level shortcomings from architectural limitations.
These remain diagnostics rather than proofs.
When computational budgets differ across architectures, sparse frontiers should also be interpreted asymmetrically: expansion beyond a densely sampled frontier provides relatively strong evidence of increased capacity, whereas an apparent shortfall may still reflect sparse exploration.

MCF also does not determine when refinement should stop.
When repeated observations of the target phenomenon are available, variation across those observations provides one natural reference point.
Once simulation discrepancies approach the variation observed across empirical instances, further improvement may increasingly reflect overfitting realization-specific differences rather than better generalization to the target phenomenon.
This empirical benchmark is not universal, however.
The fidelity required in practice depends on the simulation's intended use and on which errors matter to its users.
MCF characterizes attainable performance; whether that performance is \textit{good enough} remains a purpose-dependent validation judgment \cite{lempert2002agent}.


\paragraph{Limitations.}
Our conclusions are conditional on several choices.
For example, MCF characterizes capacity relative to a chosen boundary between architecture and configuration.
We held the language model and prompts fixed within each architecture, yet LLM outputs are sensitive to prompt wording and format~\cite{sclarQuantifyingLanguageModels2024, zhengLargeLanguageModels2024}.
Treating prompt variants as configurations would enlarge the admissible configuration space and could widen the M1 and M2 frontiers.

Second, iterative development on the same empirical data risks fitting the idiosyncrasies of one event.
Grounding each refinement in observed behavior, holding the configuration space fixed, and scoring the worse of two years reduce this risk, but a held-out event remains the decisive test. 
We therefore plan to evaluate the final architecture prospectively on the 2027 commencement.

Third, the LLM frontiers rest on 16 configurations with one seed each, so stochastic variation in the language model is not separated from configuration effects.
Finally, our evidence comes from a single venue observed through one camera and evaluated in one region; party structures were assumed from the seating layout; and the LLM layers depend on a commercial model whose behavior may change over time.
To support replication, we will release the simulator, prompts, and evaluation code.








\section{Conclusion}
\label{sec:conclusion}

Integrating LLMs into agent-based models enlarges the design space and makes it harder to distinguish configuration problems from architectural limitations.
We introduced the Model Capacity Frontier, which characterizes the empirical performance attainable by an architecture across its admissible configurations and uses frontier shifts to guide simulation development.

Applying MCF to a 12,000-agent crowd simulation, we found that a conventional egress model could not reproduce observed flow direction across the explored configurations.
An LLM decision layer expanded attainable flow performance but exposed missing post-ceremony lingering behavior; adding explicit gathering then improved temporal and spatial fit while retaining much of the flow gain.
MCF also revealed improvements that selected configuration comparisons could miss and compensating errors hidden by calibration within a single architecture.

Our case study suggests that LLM-supported ABMs should be developed by characterizing the empirical capacity of each architecture across its admissible configurations, diagnosing where that capacity remains limited, and introducing architectural refinements when further configuration is unlikely to resolve those limitations.






\bibliographystyle{ACM-Reference-Format} 
\bibliography{b}

\appendix
\setcounter{figure}{0}
\setcounter{table}{0}
\renewcommand{\thefigure}{App.\arabic{figure}}
\renewcommand{\thetable}{App.\arabic{table}}
\renewcommand{\theHfigure}{App.\arabic{figure}}
\renewcommand{\theHtable}{App.\arabic{table}}

\begin{figure}[t!]
  \centering
  \includegraphics[width=0.9\columnwidth]{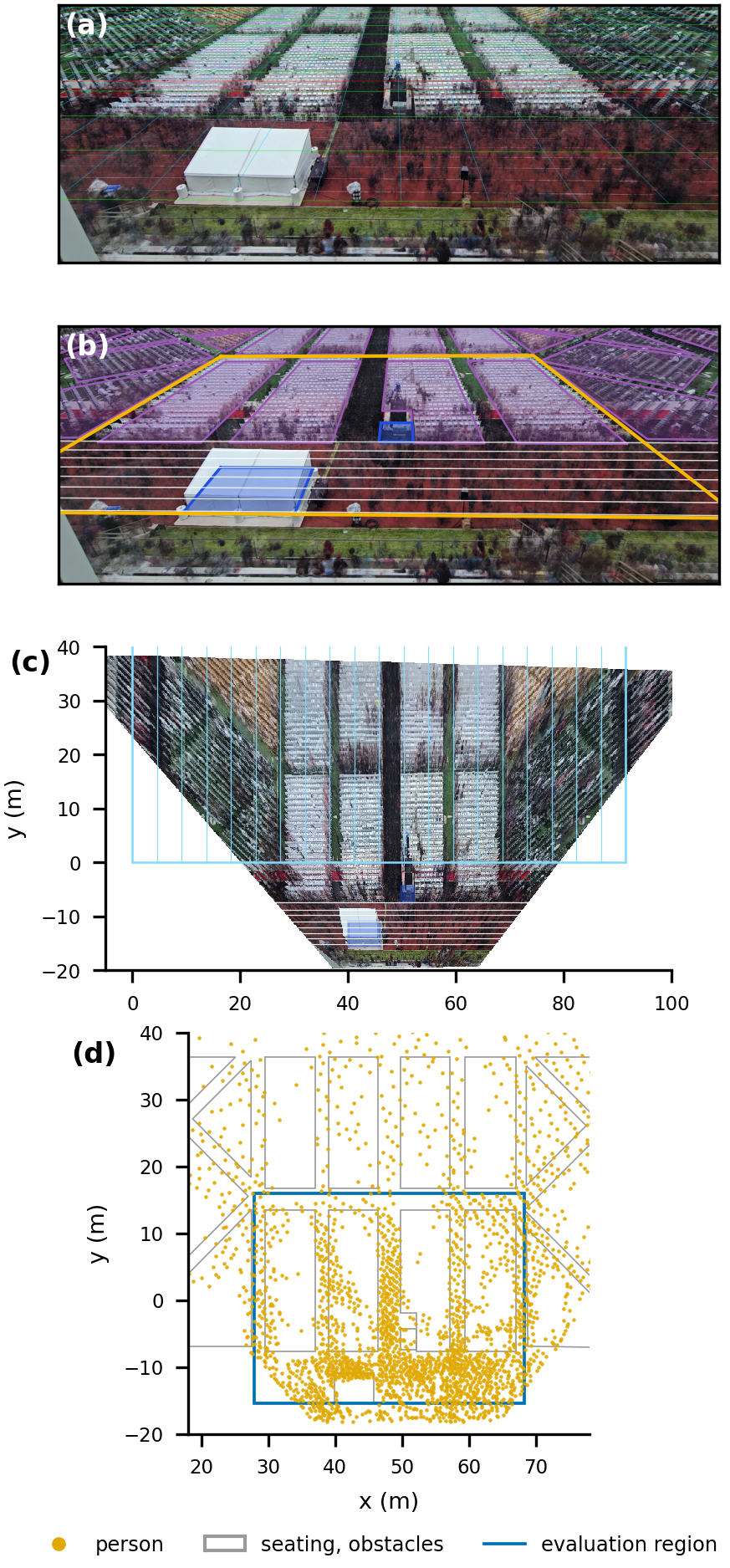}
  \caption{Calibration details from the video to the simulation's map. (a) A grid in world meters projected back onto the camera's background. (b) The simulation's arena projected onto the same view: seating (purple), obstacles (blue), running track lanes (white) and the evaluation region (yellow). (c) The camera view projected into the top-down frame of the simulation. (d) DM-Count's density map at $t = 7$\,min on the simulation map.}
  \label{fig:calib}
\end{figure}

\section{AI Usage Disclosure}

The authors acknowledge using Claude Code sessions running Opus 5/5.5 and Fable 5.1 to assist with simulation code, analysis, drafting, and figure polishing. The authors retained control over the experiments and made all design decisions.

\section{Calibration of Arena}
\label{app:calibration}

\paragraph{Determining meters in the video.} To allow the video and simulation to share one unit of measurement, we calibrate them against the dimensions of a standard football field. Between the goal lines a field measures 91.44 by 48.77\,m, and a yard line appears every 4.57\,m.

We record the commencement through a camera in the press box. For each year we map the image to the ground with a homography $H$ and a radial lens correction $k_1$, fitted to landmarks that we clicked by hand. In a wide-angle camera, straight lines on the ground appear slightly bent in the image. We apply a radial lens correction to adjust each pixel $\mathbf{p}$ from the image center $\mathbf{c}$,
\[
  \mathbf{p}' = \mathbf{c} + (\mathbf{p} - \mathbf{c})\left(1 + k_1 \lVert \mathbf{p} - \mathbf{c} \rVert^2\right).
\]
We choose $k_1$ such that running track lines become straight in the corrected image (Figure \ref{fig:calib}b). Next, we fit $H$, a homography matrix to landmarks across the image, such as the running track's lane lines and the football field's yard lines. Leaving each landmark out in turn, the model predicts their locations within a median of 0.16\,m (2025) and 0.20\,m (2026). 


\section{Measurement of the crowd}
\label{app:measurement}

\subsection{Crowd detection} We compared three counting setups on 11 image patches spread out over both years' videos, where every head of a person was labeled by hand. We also compared 2 hybrid setups, where density-based methods are applied in medium-dense areas and a conventional head detector in sparse ones. The setups are as follows:
\begin{enumerate}
    \item A YOLOv5 model ~\cite{Jocher_YOLOv5_by_Ultralytics_2022} trained on CrowdHuman~\cite{shaoCrowdHumanBenchmarkDetecting2018}, a benchmark for detecting humans in a crowd. The model is taken off-the-shelf from HuggingFace\footnote{\url{https://huggingface.co/MK-CUPIST/crowdhuman_yolov5m}}.
    \item CSRNet, a density-based method \cite{liCSRNetDilatedConvolutional2018}. Public weights are retrieved from and distributed by the LWCC library\footnote{\url{https://github.com/tersekmatija/lwcc}}.
    \item DM-Count, a second density-based method \cite{wangDistributionMatchingCrowd2020} also used through the LWCC library.
    \item (Hybrid) YOLOv5 $\times$ CSRNet
    \item (Hybrid) YOLOv5 $\times$ DM-Count
\end{enumerate}
Table \ref{tab:counting instruments} presents the results. Across the different density levels, the hybrid DM-Count setup has the lowest mean absolute at all density levels except the sparse condition. This becomes our main way to count people in the evaluation region $R$. 

\paragraph{Hybrid.} DM-Count is accurate in medium crowds but over-counts the jam by a steady factor, so we calibrate it against the hand labels in the following steps.
\begin{enumerate}
    \item \emph{Estimate the density.} DM-Count returns a number of people; dividing it by the area of the region gives an estimated density $\rho$.
    \item \emph{Measure the error in known settings.} Compare DM-Count's count with the hand count at different densities. In medium crowds (0.4 to 1.5\,persons/m$^2$, n=5), DM-Count counts slightly low, by a factor of 0.944 at a median estimate of 0.78\,persons/m$^2$. In dense crowds (above 1.5\,persons/m$^2$, n=4), it counts high, by a factor of 1.273 at a median estimate of 2.29\,persons/m$^2$.
    \item \emph{Interpolate the error for different densities.} Interpolate linearly between the points above to obtain a correction factor $r(\rho)$ for any density. Divide each DM-Count count by $r(\rho)$ at its estimated density.
\end{enumerate}
In addition to this, hybrid setups use the head detector in sparse crowds. On the two sparse patches it finds 18 of 18 and 36 of 40 people, so we divide its count by 0.95 to correct the small undercount, similarly to step 3. To prevent jumps in the count value, we take a weighted average of the two corrected counts,
so the count is from the detector below 0.15\,persons/m$^2$, DM-Count above 0.35, and a linear mix in between.

\begin{table}[h]
  \caption{Mean absolute percentage error (lower is better) of each counting setup on 11 hand-labeled image patches, sorted by crowd density (persons/m$^2$; 2, 5, and 4 images). A hybrid setup uses the head detector in sparse crowds and the network in dense ones. We settle on the DM-Count hybrid.}
  \label{tab:counting instruments}
  \small
  \setlength{\tabcolsep}{3.5pt}
  \begin{tabular}{@{}lrrrrr@{}}
    \toprule
    & head & & & CSRNet & DM-Count \\
    crowd density & detector & CSRNet & DM-Count & (hybrid) & (hybrid) \\
    \midrule
    all               & 37\% & 16\%         & 20\% & 11\% & \textbf{7\%} \\
    sparse ($<$0.3)   & 5\%  & \textbf{3\%} & 34\% & 6\%  & 13\% \\
    medium (0.3--1.5) & 21\% & 18\%         & 9\%  & 17\% & \textbf{7\%} \\
    dense ($>$1.5)    & 74\% & 19\%         & 28\% & 5\%  & \textbf{3\%} \\
    \bottomrule
  \end{tabular}
\end{table}

\subsection{Flow direction}
The flow measurement uses Farneb\"ack optical flow \cite{farnebackTwoFrameMotionEstimation2003} between video frames 0.5\,s apart. We divide the arena into 2\,m cells and sum the average direction for each cell across the entire video, counting any pixel faster than 0.3\,m/s as a walker. In the simulation, each agent's velocity is measured in the same way between two samples. Since the simulation records positions once every 10 seconds, an agent's velocity is its net movement over 10\,s. Thus, we compare only direction, not speed.

\begin{figure}[t]
  \centering
  \includegraphics[width=\columnwidth]{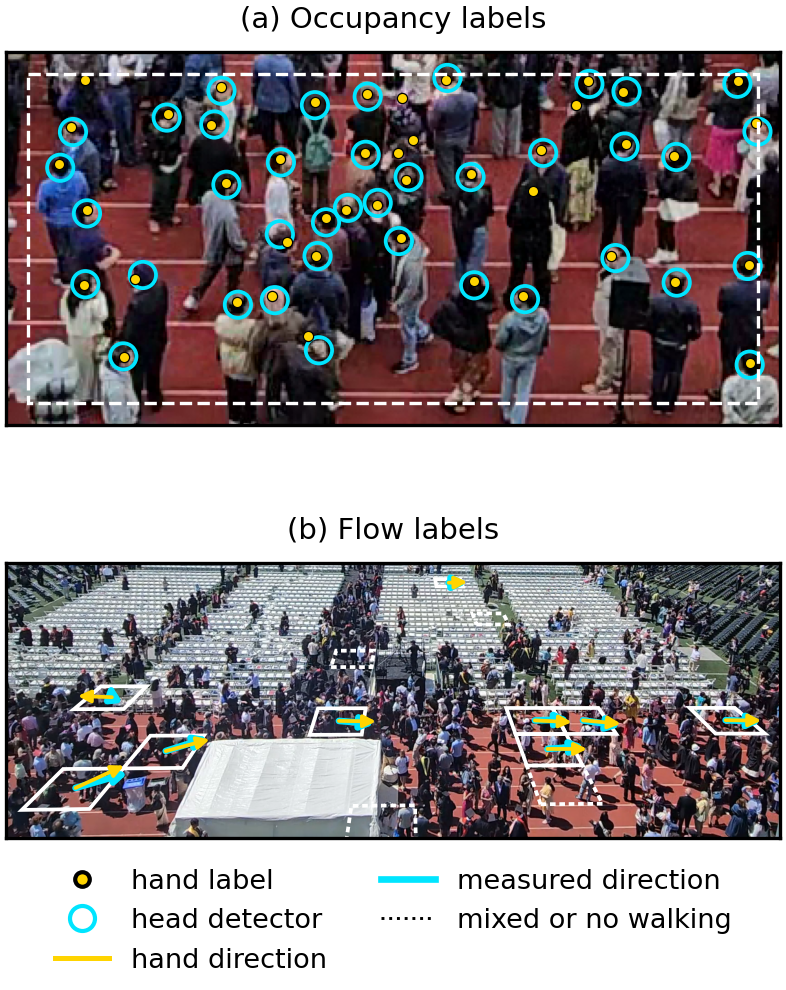}
  \caption{Hand labels for the two measurements. (a) Heads in a patch, labeled by hand (yellow dots). Cyan circles are the head detector's detections. (b) Flow direction: the 13 sampled cells of one minute. Yellow arrows are the hand labeled direction and cyan arrows are measured mean direction; dotted cells were marked as mixed or without walking.}
  \label{fig:hand_labels_flow}
\end{figure}

To check the validity of the flow detection we hand annotate six one-minute clips per year spread across the video. In each clip, a human annotator draws the direction in which most people in the cell walked (Figure \ref{fig:hand_labels_flow}). For 95 annotated cells, we measure the angle between the annotator's direction and the measured mean direction. The median difference in angle is 12$^\circ$.

\section{Walking and Routing}
\label{app:pathfinding}

\begin{figure}[t]
  \centering
  \includegraphics[width=\columnwidth]{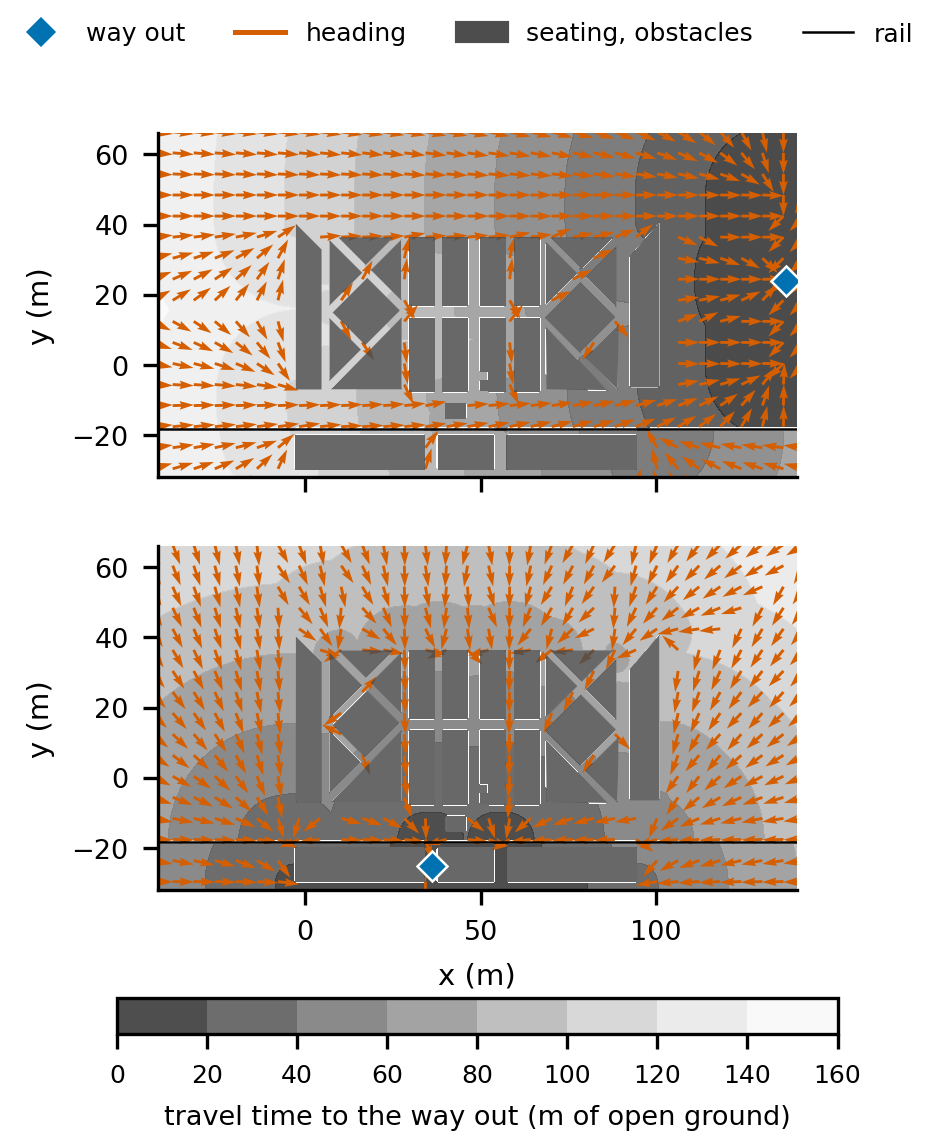}
  \caption{Navigation fields of two ways out, presented as examples. Different shading levels describe the travel time to the way out (m of open ground).}
  \label{fig:navfield}
\end{figure}

\paragraph{Walking.} 
The walker sets how fast an agent goes along the route, using a collision-free speed model~\cite{tordeuxCollisionFreeSpeedModel2016, chraibiJuPedSim2026}. An agent walks at its own desired speed $v_i^{0}$ unless someone is close ahead, in which case it slows down using
\[
  v_i = \min(v_i^{0},\ \max(0,(s_i-\ell)/\mathcal{T})),
\]
where $s_i$ is the free distance to the agent ahead, $\ell$ the desired standstill spacing, and $\mathcal{T}$ the desired time gap. 
Agents additionally steer away from nearby pedestrians using steering strength $a$ and interaction range $D$, following the formulation implemented in JuPedSim~\cite{chraibiJuPedSim2026}.

\paragraph{Routing.} Each agent needs a route to its desired destination that goes around the seating and other obstacles. Instead of individually planning a path for every agent, we precompute a \emph{navigation field} for every exit and POI in the stadium and let every agent use its direction. 

The arena is divided into cells of 0.5\,m, each assigned a traversal cost: 1 per meter on open ground, 20 inside seating and obstacles, and 500 on the rail at the foot of the grandstand, except at its four exit stairways. These values are a design choice rather than carefully calibrated; the purpose is to choose the values large enough that no route will cross the seating when it can go around, but still finite so routes can be calculated. For a destination, we then compute the travel time $\tau(\mathbf{x})$, which is the least total cost of reaching the destination from each point. It is the solution of the eikonal equation
\[
  \lVert \nabla \tau(\mathbf{x}) \rVert = c(\mathbf{x}), \qquad \tau = 0 \text{ at the destination},
\]
where $c$ is the cost per meter at $\mathbf{x}$, solved through the fast marching method \cite{sethianFastMarchingLevel1996}. 
At each step, an agent heads in the direction in which the travel time falls fastest $-\nabla\tau$ (Figure~\ref{fig:navfield}).

\paragraph{Crowd-aware routing in M1 and M2.}
In M0 the fields stay fixed for the whole run, but in M1 and M2, every 30\,s we adjust the cost of each cell according to its density. Weidmann's fundamental diagram~\cite{ weidmannTransporttechnikFussgaenger1992, kretzPedestrianTrafficQuickest2009} gives a walking-speed-density relationship
\[
  v(\rho) = v_0 \left[ 1 - \exp\!\left( -\gamma \left( \frac{1}{\rho} - \frac{1}{\rho_{\max}} \right) \right) \right],
\]
with $\gamma = 1.913$\,m$^{-2}$ and $\rho_{\max} = 5.4$\,m$^{-2}$. We cap the cost at 20 so that traffic never costs more than moving through seating obstacles.
This information is used in the prompts (Appendix \ref{app:prompts}) to describe crowdedness, allowing LLM agents to consciously choose to avoid crowded areas. 

\section{Configuration Space $\Theta$}
\label{app:config}

We describe below how the pedestrian literature and video inform our values in Table \ref{tab:theta}:
\begin{itemize}
    \item \textbf{Leaving delay scale} $\sigma_d$, 0--182: Each agent stays seated for a delay drawn from the half-normal distribution $|\mathcal{N}(0,\sigma_d^2)|$. We bound this by inspection of the video, in which most people are out of their seats within about three minutes. We allow roughly twice this duration as headroom. For a half-normal distribution, 90\% of the draws lie below $1.645\,\sigma_d$, because $P(|Z| < 1.645) = 0.90$ for a standard normal $Z$. Setting $1.645\,\sigma_d = 300$\,s gives the upper bound $\sigma_d = 300/1.645 \approx 182$\,s.
    \item \textbf{Mean desired speed} $\bar v^0$, 1.0--1.6\,m/s: This is not measurable from the video because it is the implicit desired speed on open ground, and the video only contains speeds within traffic. Instead, we set the range to cover valued reported by Davidich et al. (0.97--1.04\,m/s), Bosina et al. (1.0 to 1.6\,m/s), and Weidmann (1.34\,m/s) \cite{davidichPredictingPedestrianFlow2013, bosinaEstimatingPedestrianSpeed2017, weidmannTransporttechnikFussgaenger1992}.
    \item \textbf{Spread of desired speed} $s_{v^0}$, 0.2--0.4\,m/s: For the same reasons, we draw this range from related work, namely Weidmann (0.26\,m/s) and Buchm\"uller (0.37\,m/s) \cite{weidmannTransporttechnikFussgaenger1992, buchmullerParametersPedestriansPedestrian2006}.
    \item \textbf{Time gap} $\mathcal{T}$, 0.8--1.2\,s: Published values are from Tordeux et al. (1.0\,s) and Seyfried (1.06\,s) \cite{tordeuxCollisionFreeSpeedModel2016, seyfriedFundamentalDiagramPedestrian2005}; we allow $\pm0.2$ around 1.0 s.
    \item \textbf{Standstill spacing} $\ell$, 0.3--0.5\,m: Published values are again from Tordeux et al. (0.3\,m) and Seyfried (0.36\,m), but also a default of the JuPedSim library (0.4\,m)~\cite{tordeuxCollisionFreeSpeedModel2016, seyfriedFundamentalDiagramPedestrian2005, chraibiJuPedSim2026}
    \item \textbf{Steering strength} $a$, 0--16, and \emph{range} $D$, 0.05--0.3\,m: JuPedSim's defaults are  ($a=8$, $D=0.1$\,m), while Tordeux reports ($a=5$) \cite{chraibiJuPedSim2026, tordeuxCollisionFreeSpeedModel2016}.
\end{itemize}

\section{Power analysis}
\label{app:power}
In Section \ref{sec:development}, we discuss how only 16 configurations in M1 and M2 were evaluated due to required LLM calls. We chose 16 configurations based on a power analysis. Across the three metrics, the weakest matched-configuration effect was approximately $d_z = 0.90$. At this effect size, roughly 16 configurations give 80\% power with Bonferroni-adjusted $\alpha = 0.05/3$. 

\section{LLM Implementation Details}
\label{app:prompts}
An agent is prompted for a plan when its family's leaving time arrives, when its plan is completed, when the agent is blocked (less than 1\,m of progress in 60\,s; a stall within 8\,m of its target counts as arrival), when something changes in its family (a member arrives, the family comes together, a member sets off to leave), and otherwise every 300\,s. 

The language model is gpt-6-luna used through the API with no extended reasoning. In the case of malformed or missing responses, the agent keeps its current plan, or tries again after 5 seconds if it is its first decision. However, only 0.010\% of M1 responses and 0.006\% of M2 responses trigger the fallback.

\subsection{Messaging and Family Coordination}
Both architectures share the same mechanism for agreeing on a plan within a family. The family agrees on the place with the most proposals, and all family members adopt that action. In the event of a tie, the leader is the tiebreaker.

In M1, family members cannot vote to meet at POIs within the stadium; they can only adopt the same exit choice. In M2, family members can propose to meet at POIs such as the running track area or stage front. If agreed upon, the group gets assigned the closest point within the POI's range. Once every member is within 8\,m of that point, the members group up within 3\,m, then vote on a way out and leave together. If traffic keeps them from coming within 3\,m of each other for 180\,s, but they have all been within 8\,m, then the family counts as met. If the family has not gathered after 180\,s while more than half the members are outside of the POI (or after 540\,s in any scenario), it discards the current plan and decides a new plan.


\newsavebox{\menuMone}\newsavebox{\menuMtwo}
\begin{lrbox}{\menuMone}\begin{minipage}[t]{0.47\textwidth}\scriptsize
\begin{alltt}\textrm{\textbf{M1}}
You are one person at the end of a university
commencement in a stadium. Answer as that
person, in the first person, briefly.

[... who you are, where you are, your family,
  messages, what you have done, what you
  thought last time ...]

What you can do now (pick one id):
  wait:120: wait here for 2 minutes
  goto:stage_front: go to the front of the
    stage: the open ground in front of the
    stage, where people can stop for a few
    minutes for pictures with the stage
    behind them, then move on; about 40 m
    on foot; few people there now
  leave:bend_west: leave via the west bend
    of the track: ... into the main campus,
    where the college diploma ceremonies and
    receptions are held from 1 p.m.; about
    120 m on foot; few people on the way
  leave:stairs_east: leave via the inner east
    stairway up through the stand: ... down
    into the campus garage, the paid garage
    where guests who reserved a space for the
    day parked; about 60 m on foot; some
    people on the way
  [... other ways out ...]

propose: ... the id of one leave via line,
  meaning we all head to that exit now and
  meet beyond it, outside; or the id of one
  go to line that is a meeting place ...;
  a photo spot or a wait is never a family
  plan

Answer:
  option:  leave:bend_west
  reasoning: "The west bend is recommended
    for my section, so I'll head there and
    meet my family beyond it."
  propose: "leave:bend_west"
\end{alltt}
\end{minipage}\end{lrbox}
\begin{lrbox}{\menuMtwo}\begin{minipage}[t]{0.47\textwidth}\scriptsize
\begin{alltt}\textrm{\textbf{M2}}
You are one person at the end of a university
commencement in a stadium. Answer as that
person, in the first person, briefly.

[... who you are, where you are, your family,
  messages, what you have done, what you
  thought last time ...]

What you can do now (pick one id):
  wait:120: wait here for 2 minutes
  goto:stage_front: go to the front of the
    stage: the open ground in front of the
    stage, where people can stop for a few
    minutes for pictures with the stage
    behind them, then move on; about 40 m
    on foot; few people there now
  leave:bend_west: leave via the west bend
    of the track: ... into the main campus,
    where the college diploma ceremonies and
    receptions are held from 1 p.m.; about
    120 m on foot; few people on the way
  leave:stairs_east: leave via the inner east
    stairway up through the stand: ... down
    into the campus garage, the paid garage
    where guests who reserved a space for the
    day parked; about 70 m on foot; some
    people on the way
  [... other ways out ...]
\textbf{+ goto:near_track: go to the near track:}
\textbf{+   the running track in front of the}
\textbf{+   stand, below the rail; about 40 m on}
\textbf{+   foot; some people there now}

propose: ... the id of one leave via line,
  meaning we all head to that exit now and
  meet beyond it, outside; or the id of one
  go to line that is a meeting place ...;
  \textbf{a go to line of a photo place is a}
  \textbf{meeting place too}; a wait is never a
  family plan

Answer:
  option:  goto:near_track
  reasoning: "My family is still in the stand,
    so I'll head to the near track as a clear
    nearby meeting place."
  propose: \textbf{"goto:near_track"}
\end{alltt}
\end{minipage}\end{lrbox}
\twocolumn[{%
  \subsection{Agent action menu}
  \centering
  \fbox{\usebox{\menuMone}}\hfill\fbox{\usebox{\menuMtwo}}
  \captionof{figure}{The menu shown to one student against at its first decision (abridged) in both M1 and M2. Every prompt opens with the same one-line instruction, followed by the facts of the agent's situation, the menu, and the instruction for the answer. M2 adds the near track as a meeting place and clarifies how a family can agree to meet at a place inside the stadium (bold; the new menu lines are marked +). In M1 the graduate walks to the stage front to find its family, but it can propose only an exit as the family’s plan; in M2 it proposes the near track, and its family agrees.}
  \label{fig:menu_m1m2}
  \vspace{1.5em}
}]



\end{document}